\documentclass[11pt,fleqn]{elsart}

\usepackage{amsmath,amssymb}
\usepackage{graphicx,psfrag,here,epsfig}
\usepackage{pstricks}
\usepackage{pst-node}
\usepackage{epsf}
\usepackage{citesort}
\usepackage{exscale}
\usepackage{longtable}

\allowdisplaybreaks[1]

\newfont{\hermes}{cmtt10 at 10pt}

\newcommand{\fs}{\, . \,}
\def\co{\; \; ,}

\newcommand{\nn}{\nonumber\\[1.5ex]}

\def\bea{\begin{eqnarray}}
\def\eea{\end{eqnarray}}
\def\eq{\begin{eqnarray}}
\def\en{\end{eqnarray}}
\def\be{\begin{equation}}
\def\ee{\end{equation}}
\newcommand{\bc}{\begin{center}}
\newcommand{\ec}{\end{center}}

\renewcommand{\theequation}{\arabic{equation}}

\newcommand{\myfrac}[2]{\mbox{\large{$\frac{#1}{#2}$}}}

\newcommand{\mk}{M_K}

\newcommand{\cO}{\mathcal{O}}

\newcommand{\Lb}{{\bar{L}}}

\newcommand{\ed}{\end{document}}

\begin{document}

%\runauthor{ Gasser, Haefeli, Ivanov and  Schmid}
\renewcommand{\theequation}{\arabic{equation}}
\begin{frontmatter}

\begin{flushright}
IFIC/07-12
\end{flushright}

\title{\Large\bf Integrating out strange quarks in ChPT}

\author[Bern]{J.~Gasser},
\author[Valencia]{Ch.~Haefeli},
\author[Dubna]{M.A.~Ivanov}
\author[Bern]{and M.~Schmid}
\address[Bern]{Institute for Theoretical Physics, University of Bern,
Sidlerstr. 5, CH--3012 Bern, Switzerland}
\address[Valencia]{
        Departament de F\'{\i}sica Te\`orica,
         IFIC, Universitat de Val\`encia -- CSIC, %%%\\
         Apt. Correus 22085, E--46071 Val\`encia, Spain}
\address[Dubna]{Laboratory of Theoretical Physics,
Joint Institute for Nuclear Research, \\
141980 Dubna (Moscow region), Russia}

\begin{abstract}
We study three flavour  chiral perturbation theory in a limit where the
strange quark mass is much larger than  the external momenta and
 the up and down quark masses, and where the external fields are those of
 two--flavour chiral perturbation theory.
 In this case, the theory reduces to the one
 of $SU(2)_L\times SU(2)_R$. Through this
 reduction, one can work out the strange quark mass dependence of the LECs in
 the two--flavour case. We present the pertinent relations at two--loop order
 for  $F,B$ and $l_i$.
\end{abstract}


\end{frontmatter}

% ---- SECT 1 ----------------------------------------------------------

\noindent{\bf 1.}
We  consider Green functions of quark currents in the
framework of QCD with three flavours. 
At low energies, the Green functions  can be analysed in the 
framework of chiral perturbation theory (ChPT)\cite{weinberg,glann,glnpb}. It
is customary to perform the pertinent quark mass expansion either around
$m_u=m_d=0$, with the strange quark mass held fixed at its physical value
(ChPT$_2$), or to consider an expansion in all three quark masses, around
$m_u=m_d=m_s=0$ (ChPT$_3$).
The relevant  effective lagrangians contain  low--energy constants (LECs)
which are not determined by chiral symmetry alone. The two expansions are not
independent:  
one can express the LECs in the
two--flavour case through the ones in ChPT$_3$. These relations were given at
one--loop order in \cite{glnpb} and were used to obtain information on the LECs
in ChPT$_3$ from those known  in the two--flavour case. 
Because there are many two--loop calculations available now, 
both in the two-- and three--flavour case \cite{Bijnens:2006zp}, 
it is expedient to have the relevant relations between
the LECs at two--loop accuracy as well, both, 
to obtain more additional information, 
 and for internal consistency checks. It is the purpose 
of this letter to provide the relations  
that occur  at order $p^2$ and $p^4$ in ChPT$_2$.

We comment on related work that is available in the literature. i) The
strange quark mass dependence of the
ChPT$_2$ LECs at order $p^2,p^4$  can be worked out at two--loop order from existing
two--loop calculations in the three-flavour sector, see below. ii) The strange quark
mass expansion of the ChPT$_2$ LEC $B$ ($F^2B$) was already  provided at this
accuracy in 
Ref.~\cite{Kaiser:2006uv} (\cite{Mouss:Sigma}). 
 iii) The authors of Refs.~\cite{GChPT,dcpipikk} 
investigate what happens
 if chiral symmetry breaking exhibits different patterns in ChPT$_2$ 
and ChPT$_3$. The literature  on the subject 
may be traced from Ref.~\cite{dcpipikk}.
 In this scenario, a substantial strange quark mass dependence may 
show up, as a result of which ChPT$_3$ must be reordered
 and the effect of vacuum fluctuations of $\bar ss$ pairs 
 summed up.   
 Whether the relations provided below
favour such a situation is not investigated here -- 
 the present work just 
provides the {\it algebraic} dependences of the ChPT$_2$ LECs on the
 strange quark mass, at two--loop order.
iv) Analogous work was performed at one--loop accuracy in the baryon sector
in Ref.~\cite{meissner_frink}, and for electromagnetic corrections in
 Refs.~\cite{gasser_rusetsky,jallouli_sazdjian,NehmePhD}.

\vskip2mm

% ---- SECT 2 ----------------------------------------------------------

\noindent{\bf 2.}
We first illustrate how  the relations  between the LECs emerge, and
 consider the pion matrix element of the vector current,
 \begin{equation}
\langle \pi^+(p')\,|\tfrac{1}{2}(\bar u {\gamma_\mu} u-\bar d{\gamma_\mu}
d)|\pi^+(p)\rangle = (p+p')_\mu 
F_V(t)\,\,;\,\,
t=(p'-p)^2\,,   
 \end{equation}
in the chiral limit $m_u=m_d=0$. In the three--flavour case, at
one--loop order, the result reads in $d$ space--time dimensions
\begin{equation}
  \label{eq:loopsu3}
F_{V,3}(t)
=1+\frac{t}{F_0^2}\left[\Phi(t,0;d)+\tfrac{1}{2}\Phi(t,M_K;d)\right] 
+ \frac{2L_9 t}{F_0^2}\,.  
\end{equation}
The loop function $\Phi$, generated by pions and kaons running in the
loop, is given by
\begin{equation}
\Phi(t,M;d)=\frac{\Gamma(2-\frac{d}{2})}{2(4\pi)^{d/2}}\int_0^1
du \,u^2\left[M^2-\myfrac{t}{4}(1-u^2)\right]^{\frac{d-4}{2}}\,.  
\end{equation}
Furthermore,  $F_0$ denotes the pion decay constant
at $m_u=m_d=m_s=0$, and $L_9$ is one of the LECs in ChPT$_3$ at order $p^4$.

In ChPT$_2$, the corresponding one--loop expression is
\begin{equation}
F_{V,2}(t)=1+\frac{t}{F^2} \Phi(t,0;d)-\frac{l_6t}{F^2}\,\,,  
\end{equation}
where $F$ denotes the pion decay constant at $m_u=m_d=0,m_s\neq 0$, and where
$l_6$ stands for a low--energy constant in ChPT$_2$  at order $p^4$. 
If one identifies $F$ with $F_0$ at this order, the  expressions $F_{V,3}$ and
$F_{V,2}$ still differ in the
coefficient of the term proportional to $t$, and in the contribution
$\Phi(t,M_K;d)$, which is absent in the two--flavour case, because kaons are
integrated out in that framework.

\begin{figure}
\begin{center}
\epsfig{file=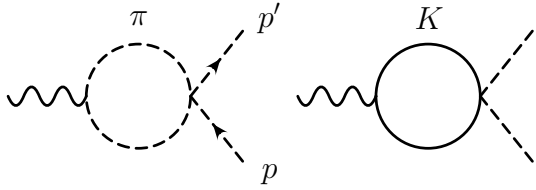,width=7cm}
\end{center}
\caption{ Loops that generate branch points in the function
  $\Phi(t,M;d)$. Dashed (solid) lines denote pions (kaons), the wiggly line is
  the vector current.
Tadpoles are not displayed -- these contribute with a constant
  term to $F_V$, as a result of which the form factor is 
normalised to one at $t=0$, as is required
  by the pertinent Ward identity.}\label{fig:branch}
\end{figure}

To proceed, we note that the loop function $\Phi$ 
is holomorphic in the complex $t$--plane, cut
along the real axis for Re $t\geq 4M^2$. We display in Fig.~\ref{fig:branch}
the loops that generate these branch points: pions (kaons) for the one at
$t=0\,\, (t=4 M_K^2)$. Therefore,
$\Phi(t,0;d)$ has a branch point at $t=0$, whereas $\Phi(t,M_K;d)$
reduces to a polynomial
at  $t/M_K^2\ll 1$,
\bc
\ec
\begin{equation}
  \label{eq:polynom}
\Phi(t,M_K;d)=\sum_{l=0}^{\infty} 
\Phi_l(M_K,d)\left(\frac{t}{M_K^2}\right)^l\,.  
\end{equation}
Let us discard for a moment 
the terms of order $t$ and higher in this expansion.
It is then seen that
$F_{V,3}$ reduces to $F_{V,2}$, provided that we set
\begin{equation}
  \label{eq:l6L9}
l_6=-2L_9-\frac{1}{2}\Phi_0( M_K,d)\,.  
\end{equation}
At $d=4$, this
relation  reduces to the one between the 
renormalised LECs $l_6^r$ and $L_9^r$, provided in
\cite{glnpb}, see also below.

We conclude that, at low energies, the expression of the vector form factor in
ChPT$_3$ reduces to the one in the two--flavour case, up to polynomial terms of
order $t^2$ and higher, using properly matched LECs in the two--flavour
framework. A similar statement holds true for  all Green functions of quark
currents built from up and down quarks alone, see below.

% ---- SECT 3 ----------------------------------------------------------

\noindent{\bf 3.}
We  now come back to  the higher--order terms in
Eq.~(\ref{eq:polynom}).
  We start with the observation that
the term of order $t^l$ contributes at order $t^{l+1}$ to $F_{V,2}$ -- those
with $l\geq 1$ are thus of the same chiral order in $F_{V,2}$ 
as the ones generated by graphs with $l+1$ loops in ChPT$_2$.
 Apparently, one runs into a problem with power counting here: the low--energy
 expansion of the one--loop contribution in ChPT$_3$ amounts to terms of
 arbitrarily high orders in the $SU(2)\times SU(2)$ 
 expansion of $F_{V,2}$. Indeed, this is
 a rule rather than an exception: Because the strange quark mass is counted
 as a quantity of chiral order zero in ChPT$_2$, the counting of a quantity like
$t/M_K^2$ is different in the two theories.  As a result of this, 
higher--order loops in ChPT$_3$ in general start to contribute already at 
leading order in ChPT$_2$.

 As will be discussed in our forthcoming
 publication \cite{matchingII}, a   systematic and coherent
 scheme is obtained by counting
 $n$--loop contributions -- and, in particular the 
relevant LECs -- to be of order $\hbar^n$, and the strange quark mass to
 be of order $\hbar^{-1}$.  In this manner, it is easily seen that e.g. 
the term
 with $l=1$ in (\ref{eq:polynom}) contributes at order $\hbar^2$ to $F_{V,2}$
 and amounts to  a contribution from  the LECs
at two--loop order in ChPT$_2$, etc.

We call in the following the
contributions from tree, one--loop and two--loop graphs  in ChPT$_3$ 
to the two--flavour LECs {\it leading, next--to--leading 
and next--to--next--to--leading 
order} contributions, respectively. The  relation for the renormalised 
LEC $l_6^r$ is 
\begin{equation}\label{eq:l6r}
l_6^r=\alpha+\beta\,y  +O(y^2)
\co
\quad
y=\frac{B_0m_s}{16\pi^2F_0^2}
\co
\end{equation}
with coefficients $\alpha,\,
\beta,\ldots$ that are polynomials in the LECs of ChPT$_3$ and in the
logarithms of the meson 
masses\footnote{We use the following notation.
Order $p^2;p^4: F,B;l_1,\ldots , l_7, h_1,h_2,h_3$ (ChPT$_2$ \cite{glann});
$F_0,B_0;L_1,\ldots, L_{10},H_1,H_2$
(ChPT$_3$ \cite{glnpb})\,. Order $p^6:
C_1,\ldots, C_{94}$ (ChPT$_3$ \cite{bcelag,bceren}), and similarly for the
renormalised quantities $l_i^r,\ldots,C_i^r$. }.
 The quantity $\alpha$ ($\beta\,y$) denotes the NLO (NNLO) term,  generated
 by one--loop (two--loop) graphs in ChPT$_3$. 
They
are all of order $\hbar$ according to the above counting rules. [There is no
contribution from  tree graphs for $l_6^r$.]

 For illustration, we note that from Eq.~(\ref{eq:l6L9}), one finds
\begin{equation}
\alpha=\frac{1}{192\pi^2}\left[\ln (B_0m_s/\mu^2)+1\right]-2L_9^r(\mu)\,,
\end{equation}
where $\mu$ denotes the renormalisation scale.

In conclusion, the above
counting of loops and of $m_s$ in powers of $\hbar$ generates a systematic
(Laurent)--series of the LECs in the variable $y$, modulo logarithmic terms,
and greatly simplifies the counting.

% ---- SECT 4 ----------------------------------------------------------

\vskip2mm
\noindent{\bf 4.}
We now investigate the expansion of the LECs 
in general, and discuss  how one can determine the NNLO 
terms that we are after here.
 An obvious procedure
is the one  used above for $l_6^r$: One compares matrix elements, evaluated in
$SU(2)\times SU(2)$ with the same ones evaluated in $SU(3)\times SU(3)$.
In the case of $F,B, l_1,$$\ldots l_6$, available loop calculations
 allow one 
to perform the matching at NNLO. We refer to this framework as {\it method I}
 in the following.
 In order to match
the LECs at order $p^6$  to the same accuracy in this manner, 
a tremendous amount of two--loop
calculations in  ChPT$_{2,3}$ would be required.
We believe that it is fair to say that these will never be performed.

Therefore, and in order to have an independent check on the results, we
have developed \cite{mschmidphd,matchingII} a generic {\it method II},
 based on the path integral formulation of ChPT. It consists in
 the evaluation of  the local terms that are generated in the framework of
 ChPT$_3$ by graphs where heavy particles are running in the
 loops, and identifying these with local contributions generated by the
 counterterms in ChPT$_2$. This method allows one to perform the matching for
 the LECs at order $p^6$  as well. Because the 
 technique is rather involved, we defer a
 detailed description of the framework to a forthcoming  publication
 \cite{matchingII}.

\begin{table}[t]
\begin{center}
\begin{tabular}{ccccccc}
\hline\hline
LEC
&Source
&Ref.
&\hspace*{0.2cm}
&LEC
&Source
&Ref.
\\ 
\hline
$F$
&$F_\pi$
&\cite{Amoros:1999dp}
&
&
$F^2B$
&$\langle 0|\bar uu|0\rangle$
&\cite{Bijnens:2006ve,Mouss:Sigma}
\\
$l^r_{1,2}$
&$\pi\pi\to\pi\pi$
&\cite{Bijnens:2004eu}
&
&
$l_3^r$
&$M_\pi$
&\cite{Amoros:1999dp}
\\
$l_4^r$
&$F_\pi$, $M_\pi$
&\cite{Amoros:1999dp}
&
&
$l_5^r$
&$\langle 0|A^i_\mu A^k_\nu|0\rangle$\, ,
$\langle 0|V^i_\mu V^k_\nu|0\rangle$
&\cite{Amoros:1999dp}
\\
$l_6^r$
&$F_V(t)$
&\cite{Bijnens:2002hp}
&
&
$h_1^r$
&$\langle 0|\bar uu|0\rangle$ ,
$M_\pi$ , $F_\pi$
&\cite{Bijnens:2006ve,Amoros:1999dp}
\\
$h_2^r$
&$\langle 0|V^i_\mu V^k_\nu|0\rangle$
&\cite{Amoros:1999dp}
&
&
$h_3$
&$\langle 0|S^i S^k|0 \rangle$ ,
$B$
&\cite{B:h3,Amoros:1999dp}
\\[1ex]
\hline\hline\\[-1ex]
\end{tabular}
\caption{The quantities used to match  $F,B;l_i,h_i$ at NNLO 
(method I). 
  The results for
  $F,B,l_1,\ldots, l_6$ and $h_i$ agree
  with  method II \cite{matchingII},
and the result for $B$ agrees with the calculation performed 
in \cite{Kaiser:2006uv}. 
The matching for $l_7$ was only performed with method II, 
see text.} 
\label{tab:matching} 
\end{center}
\end{table}

We have performed 
the matching of $F,B;l_1,\ldots,l_6, h_i$ at NNLO in both frameworks, 
see table \ref{tab:matching} for the quantities invoked in method I.
(The analytical formulae for the two--loop ChPT$_3$ quantities are
provided in \cite{Bijnens:homepage}, whereas the ones of ChPT$_2$ are only
needed at one--loop order.).
 The results of the two calculations fully agree. This is a
highly nontrivial check on our calculation (and on the corresponding two--loop
ChPT$_3$ one). As for $l_7$, the
correlator $\langle 0| T \bar u\gamma_5 u (x) \bar u\gamma_5 u (0) |0\rangle$
might be invoked in method I, in order to check the result obtained 
with method II. This correlator is,  
however, not available in the literature to the
best of our knowledge. We found that the available 
two--loop expressions for the neutral pion
mass \cite{Amoros:2001cp} -- 
 from where one could determine $l_7$ as well in principle -- are
 too voluminous to be used for this purpose. 
 On the other hand, we have checked several 
 expressions in $l_7$ by comparing with the corresponding  
contributions  to the two-point function 
of two pseudoscalar isoscalar densities,  
using Eq.~(12.10) in Ref.~\cite{glann}.

This completes the discussion of the methods used for the determination 
of the NNLO terms in the $m_s$ expansion of the LECs. 
In the remaining part of this letter, we display the NNLO 
results for the $p^2,p^4$ LECs, and then illustrate the use of these in one
particular application. To keep this article reasonably short,
we do not display the expressions for the 
contact terms $h_{1,2,3}$. They are available from the authors upon
request, and will in addition be presented in \cite{matchingII}.

% ---- SECT 5 ----------------------------------------------------------

\noindent{\bf 5.}
All the relations may be put in a form similar to (\ref{eq:l6r}).
To render the formulae more compact, we found it convenient to 
 slightly reorder the  expansions, such that they become 
a series in  the quantity $\mk^2$,
which stands for the one--loop expression of the (kaonmass)$^2$ in the limit
$m_u=m_d=0$, see e.g. \cite{glnpb}. The result is
\begin{eqnarray}
  \label{eq:1}
Y &=& Y_0\,\left[ 1 + a_Y\,x + b_Y\,x^2 + \cO(x^3) \right]
\co
\qquad
Y = F \, , \Sigma 
\co
\nn
l^r_i &=& a_i + x\,b_i + \cO(x^2) \, ,i\neq 7\co\\[1.5ex]
l_7&=&\frac{F_0^2}{8B_0m_s}+a_7+x\,b_7+\cO(x^2)\co\nonumber\\[1.5ex]
x &=& \frac{\mk^2}{N F_0^2} \, ,\qquad 
N = 16\pi^2 \, ,
\qquad
\Sigma = F^2 B \co
\qquad
\Sigma_0 = F_0^2 B_0 \fs
\nonumber
\end{eqnarray}
We
denote the contributions proportional to $a_i\,\,\,(b_i)$ as NLO\,\,(NNLO) terms.
Note that $l_7$ receives a contribution at leading
order (LO) as well, proportional to $m_s^{-1}$. 
 The LO and NLO terms were already determined in \cite{glnpb} -- for 
convenience, we reproduce the $a_i$  here,
\begin{eqnarray}
\label{eq:2}
a_F &=&
       - 1/2\,\ell_K
       + 8\,L^r_{4}\,N
\co
\hspace*{25mm}
a_\Sigma =
       - \ell_K
       - 2/9\,\ell_\eta
       + 32\,L^r_{6}\,N
\co
\nn
a_{1} &=&
       - 1/24\,\nu_K
       + 4\,L^r_{1}
       + 2\,L_{3}
\co
\hspace*{15.5mm}
a_{2} =
       - 1/12\,\nu_K
       + 4\,L^r_{2}
\co
\nn
a_{3} &=&
       - 1/18\,\nu_\eta
       - 8\,L^r_{4}
       - 4\,L^r_{5}
       + 16\,L^r_{6}
       + 8\,L^r_{8}
\co
\nn
a_{4} &=&
       - 1/2\,\nu_K
       + 8\,L^r_{4}
       + 4\,L^r_{5}
\co
\hspace*{17.5mm}
a_{5} =
         1/12\,\nu_K
       + L^r_{10}
\co
\nn
a_{6} &=&
         1/6\,\nu_K
       - 2\,L^r_{9}
\co
\nn
a_{7} &=&
       - 5/18\,N^{-1}
       + 1/2\,\nu_K
       + 5/9\,\nu_\eta
       + 4\,L^r_{4}
       - 4\,L^r_{6}
       - 36\,L_{7}
       - 12\,L^r_{8}
\co
\end{eqnarray}
with the abbreviations
\begin{equation}
\nu_P = \myfrac{1}{2N}\left(\ell_P + 1\right) 
\co\,\,
\ell_P = 
\ln(M_P^2/\mu^2)
\co
\qquad
P = K,\eta
\fs  
\end{equation}
In analogy to $\mk$, the quantity $M_\eta$ denotes the eta mass at one--loop
order, in the limit $m_u=m_d=0$ \cite{glnpb}. 

The NNLO contributions $b_i$ are more involved. As alluded above, its
dependence on the strange quark mass only shows up logarithmically
\begin{equation}
  \label{eq:p0p1p2}
  b = p_0 + p_1\,\ell_K + p_2\,\ell_K^2   
\fs
\end{equation}
[We suppress the index $i$ of $b_i$ for convenience]. The polynomials
$p_j$ are independent of the strange quark mass and their scale dependence is
such that in combination with the logarithms it adds up to the scale
independent quantity $b$. This not only allows for a consistency
check on our calculations, but moreover offers the opportunity to rewrite $b$
 in a particular compact manner,
 \begin{equation}
   \label{eq:lnLamb}
  b = k 
  + p_2\,\ln^2 (\mk^2/\Lambda^2)
\co
\qquad
k =  
p_0 - \frac{p_1^2}{4p_2} 
\fs   
 \end{equation}
By construction the squared logarithm is scale
independent and so is the combination $k$. The explicit results for the
polynomials $p_j$ as well as the logarithm $\ln(\mk^2/\Lambda^2)$ are
displayed in Tab.[\ref{tab:1}--\ref{tab:4}]. We introduced 
scale independent LECs $\Lb_i$, 
\begin{equation}
  L_i^r = \myfrac{\Gamma_i}{2N}\left(\Lb_i + \ell_K \right) 
  \co  
\end{equation}
with $\Gamma_i$ the $\beta$ function of $L_i^r$ \cite{glnpb}, as well as 
abbreviations for  the Clausen function evaluated at
two different arguments,
\begin{eqnarray}
\rho_{1} &=& \sqrt{2}\,\mathrm{Cl}_2(\arccos(1/3)) 
\cong 1.41602
\co \quad
\rho_{2} = \sqrt{3}\,\mathrm{Cl}_2(\pi/3)
\cong 1.75793 \,,\nonumber\\[1ex]
\mathrm{Cl}_2(\theta)&=&-\frac{1}{2}\int_0^\theta
 d\phi \,\, \ln\,(4\sin^2{\frac{\phi}{2}})\fs
\end{eqnarray}

% ---- SECT 6: Numerics ------------------------------------------------

\noindent{\bf 6.} 
As an application, we discuss the strange quark mass
dependence of the scale independent LEC $\bar{l}_2$, 
\begin{equation}
  \bar{l}_2 = {3N}l_2^r(\mu) - \ln \myfrac{M_\pi^2}{\mu^2} \,  \co
\end{equation}
with $M_\pi = 139.57\mathrm{\,MeV}$. Considering
$\bar l_2$ is of interest,
 because in Ref.\cite{Colangelo:2001df} 
it has been determined from a dispersive analysis
to rather high precision,
\begin{equation}
  \label{eq:l2bar}
  \bar{l}_2 = 4.3 \pm 0.1
  \fs
\end{equation}
 One then expects that in combination with
the formulae presented here,  $\bar l_2$ provides additional
constraints on the pertinent combination of 
 three--flavour LECs. This is furthermore  supported from the
observation that -- aside from 
 the  combination $2C_{13}^r-C_{11}^r$ --  only
the two three--flavour LECs $L_2^r$ and $L_3$ appear in the analytical
expression. According to Ref.~\cite{Amoros:2001cp}, these are 
known rather precisely,
\begin{align}
\label{eq:p4LECs}
L_2^r
&
=(+0.73\pm 0.12)\,10^{-3} 
\co
&
L_3
&
=(-2.35\pm 0.37)\,10^{-3}
\fs
\end{align}
Here and in the following, the running scale is taken at $\mu=M_\rho=770
\mathrm{\,MeV}$. 
Further, at the accuracy we are working, we may identify $F_0$ with 
$F_\pi = 92.4 \mathrm{\,MeV}$. 

\begin{figure}[t]
\begin{minipage}[]{1.0\linewidth}
\epsfig{file=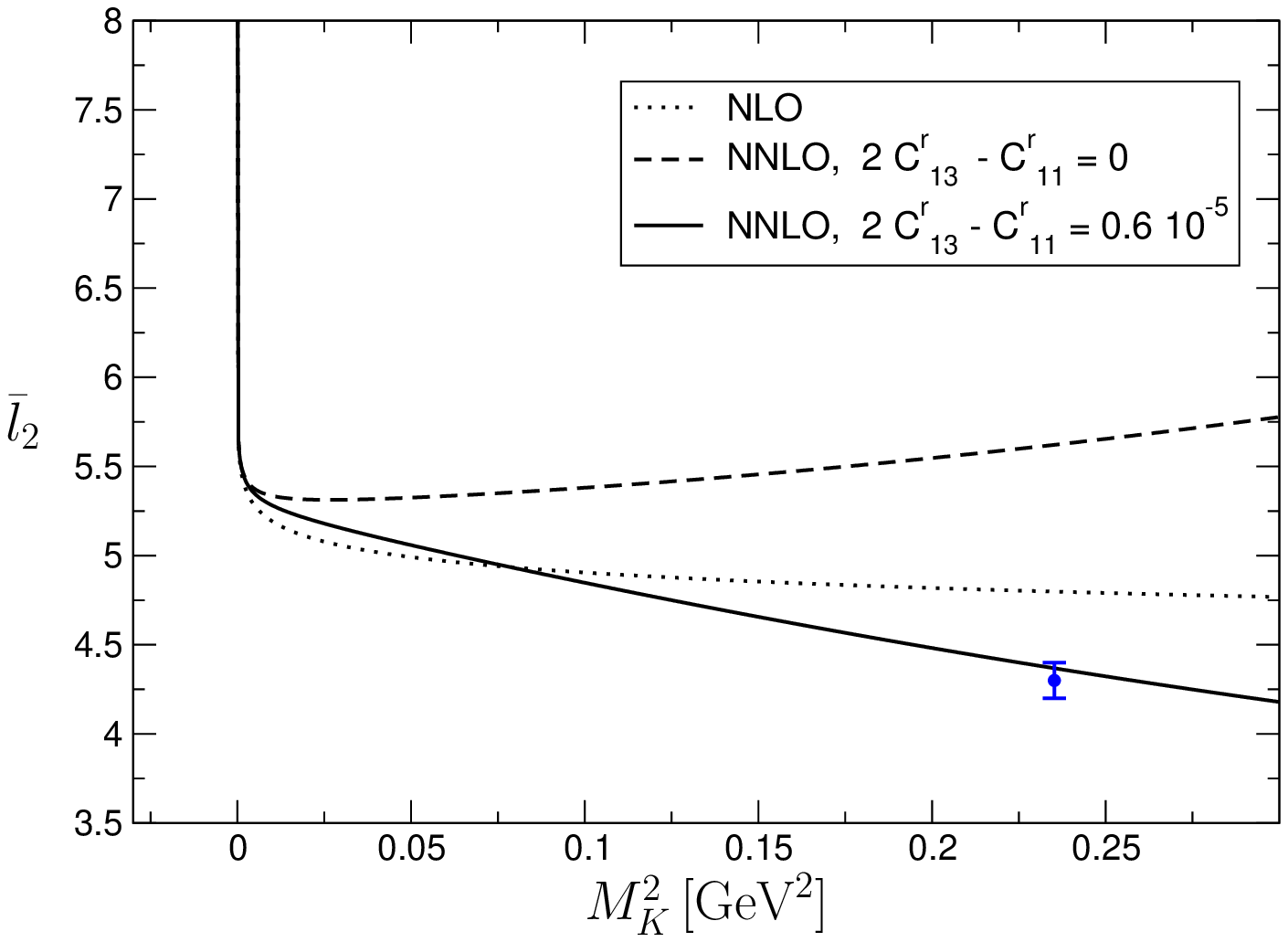,width=0.48\linewidth}
\hfill
\epsfig{file=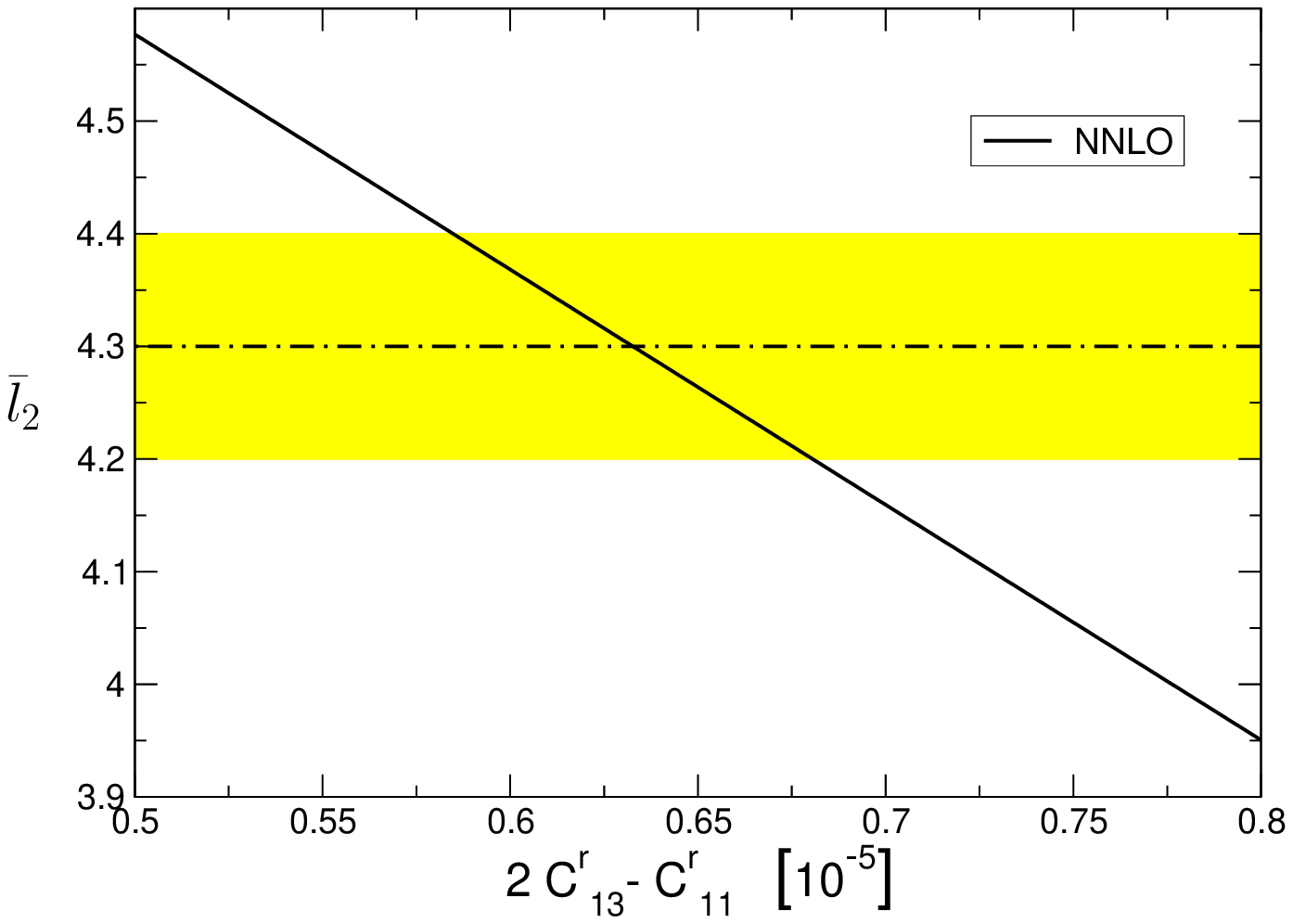,width=0.48\linewidth}\\
\end{minipage}
\caption{Left panel: Strange quark mass dependence of $\bar{l}_2$. As
  mentioned in the text, $M_K$ denotes the
  kaon mass at one--loop accuracy in the limit
  $m_u=m_d=0$. The physical value of $m_s$ corresponds to 
  $M_K \approx 485\mathrm{\,MeV}$.
  We show the NLO (dotted line) as well as the NNLO result with two
  choices for $C_{11,13}^r$: The dashed line 
  corresponds to $2C^r_{13}-C^r_{11}=0$, while the solid line is evaluated at 
  (\ref{eq:C1113}), 
  which reproduces the prediction from the dispersive
  analysis \cite{Colangelo:2001df}, Eq.~(\ref{eq:l2bar}) (data point with
  small error bar). 
  Right panel: Dependence of $\bar{l}_2$ on the $p^6$ LECs
  $C_j^r$ at the physical value of $m_s$. The dashed--dotted line with
  the error band corresponds to the data point and its error bar in the
  left panel. The running scale is taken at
  $\mu=M_\rho=770\mathrm{\,MeV}$.}\label{fig:l2.Mk2.eps}       
\end{figure}

 We illustrate the strange quark
mass dependence of $\bar{l}_2$  in Fig.\ref{fig:l2.Mk2.eps}
 (left panel), where $\bar l_2$ is shown as a function of $M_K^2$
 at $m_u=m_d=0$ (as introduced above).
The dotted line stands for the NLO approximation, and
the NNLO result is shown for two choices for  $C_j^r$: 
the dashed line displays the case $2C_{13}^r-C_{11}^r=0$, while
the solid line is worked out at 
\begin{align}\label{eq:C1113}
2C_{13}^r-C_{11}^r=0.6\cdot 10^{-5}\,,
\end{align}
chosen such that at the physical value of the strange quark 
mass, the LEC $\bar l_2$ agrees with the measured one, 
within the uncertainties. [Note  that the singular behaviour 
at $m_s \rightarrow 0$, generated
by a chiral logarithm, is not in the validity domain of our formulae any more:
the expansion performed here requires that all external 
momenta  are much smaller 
than $m_s$. Remarkably, the pertinent logarithm becomes dominant
numerically only for very small $m_s$.]

We shortly comment on the $C_{11,13}^r$  that occur in this application. In 
Ref.~\cite{Bijnens:2003xg}, $C_{13}^r$ is worked out from an 
analysis of scalar form factors. While the result is of the order found in
(\ref{eq:C1113}), its precise value
depends considerably on the input used, see table 2 in 
Ref.~\cite{Bijnens:2003xg} for more information.
In Ref.~\cite[table 12]{ResCi} estimates for both LECs $C_{11,13}^r$ are 
provided: the authors find that these  do not receive a contribution from
resonance exchange at leading order in large $N_C$ and therefore vanish at this order of
accuracy. Because the scale at which this happens is not fixed a priori, that
observation is not necessarily in contradiction with the above result \cite{matchingII}.

The impact of these LECs  on $\bar{l}_2$ is rather enhanced at physical
strange quark masses. This is illustrated in Fig.\ref{fig:l2.Mk2.eps}
(right panel). Taking the LECs $L_j^r$ from Eq.~(\ref{eq:p4LECs}) at face value,
the window for a possible choice of the $C_j^r$ is then very narrow to be in
agreement with Eq.~(\ref{eq:l2bar}).
To pin down the $C_j^r$ to good precision
including an error analysis requires
however a more thorough exploration. In particular, one has  to take into
account that in the fits performed in Ref.~\cite{Amoros:2001cp}, an estimate of order
$p^6$ counterterm contributions  was already used. This work is in progress and is deferred
to a forthcoming  publication \cite{matchingII}.
\bc
\ec

% ---- SECT 7 ----------------------------------------------------------
\enlargethispage{\baselineskip}

\noindent{\bf 7.}
In summary, we have worked out the strange quark mass dependence of the
two--flavour LECs at order $p^2$ and $p^4$. The calculation is  performed 
  at next-to-next-to leading order in
 an expansion in the quantity  $x =\mk^2/16\pi^2F_0^2$.
The result amounts to twelve  relations between the LECs in ChPT$_2$ and 
the ones in ChPT$_3$.
 Details  of the calculation will be given elsewhere \cite{matchingII}.

We have  illustrated
the use of our results  in the case of
  $\bar l_2$: its precise knowledge, together
with the known values of $L_2^r, L_3$, in principle allow one to
pin down the combination $C_{11}-2C_{13}$  rather precisely.
  Work on analogous constraints generated by the
remaining LECs at order $p^2$ and $p^4$ is in progress \cite{matchingII}.
 It will be interesting
to  merge these constraints with information on the $p^6$ LECs from
other sources \cite{Amoros:1999dp,Bijnens:2006zp,CrSource,ResCi}.

The strange quark mass dependence of the  $SU(2)$ LECs at order $p^6$ 
can be established along similar lines \cite{mschmidphd,matchingII}.
 The major burden of
 the calculation is the algebraic complexity of ChPT at $\cO(p^6)$. Various
steps and considerable progress towards 
this goal have already been made \cite{mschmidphd}.
We hope to report on the complete calculation at a later stage.

% ---- TABLES  ----------------------------------------------------------

\renewcommand{\arraystretch}{1.5}
\newcommand{\extraline}{$\\ & $}
\setlength{\LTcapwidth}{\textwidth}

\begin{longtable}{cl}
\hline\hline
& $
p_0
$ \\
\hline\\[-2ex]
\endhead
\\[-2ex]
\hline\hline
\caption[]{\rule{0cm}{2em} The polynomial $p_0$ defined in
  Eq.(\ref{eq:p0p1p2}) for $F$, $\Sigma$, $l_1^r\ldots l_6^r,\;l_7$.}
\label{tab:1}
\endfoot
$ F 
$ & $
       - 73/32
          + 2/3\,\rho_{1}

          + 1/3\,\ln\myfrac{4}{3} 

       + N \, [
          - 52/9\,L^r_{2}
          - 43/27\,L_{3}
          ]
\extraline
       + N^2 \, [
            96\,(L^r_{4})^2
          + 64\,L^r_{4}\,L^r_{5}
          - 256\,L^r_{4}\,L^r_{6}
          - 128\,L^r_{4}\,L^r_{8}
          + 32\,C^r_{16}
          ]
\extraline
       + \ln\myfrac{4}{3}\,N \, [
            128/9\,L^r_{1}
          + 32/9\,L^r_{2}
          + 32/9\,L_{3}
          - 32/3\,L^r_{4}
          ]

$ \\[1ex]
$ \Sigma
$ & $
         26/81\,\ln^2\myfrac{4}{3} 

       + N^2 \, [
            512\,L^r_{4}\,L^r_{6}
          + 256\,L^r_{5}\,L^r_{6}
          - 1024\,(L^r_{6})^2
          - 512\,L^r_{6}\,L^r_{8}
\extraline
          + 64\,C^r_{20}
          + 192\,C^r_{21}
          ]

       + \ln\myfrac{4}{3}\,N \, [
            160/9\,L^r_{4}
          + 16/3\,L^r_{5}
          - 448/9\,L^r_{6}
          + 64/3\,L_{7}
\extraline
          - 32/9\,L^r_{8}
          ]
$ \\[1ex]
1 
& $
         N^{-1} \, [
          - 73/288
          + 1/24\,\ln\myfrac{4}{3}
          + 1/16\,\rho_{1}
          ]

       + 8\,L^r_{1}
          + 2\,L_{3}
          - 4\,L^r_{4}

\extraline

       + N \, [
            8\,C^r_{6}
          - 8\,C^r_{11}
          + 32\,C^r_{13}
          ]
$ \\[1ex]
2
& $
         N^{-1} \, [
            433/288
          - 1/24\,\ln\myfrac{4}{3}
          + 1/16\,\rho_{1}
          ]

       + N \, [
            16\,C^r_{11}
          - 32\,C^r_{13}
          ]
$ \\[1ex]
3
& $
         N^{-1} \, [
            1075/2592
          - 79/288\,\rho_{1}
          ]

          - 383/1296\,\ln\myfrac{4}{3}\,N^{-1}

          + 5/81\,\ln^2\myfrac{4}{3}\,N^{-1} 

\extraline
       + N \, [
          - 128\,(L^r_{4})^2
          - 64\,L^r_{4}\,L^r_{5}
          + 768\,L^r_{4}\,L^r_{6}
          + 256\,L^r_{4}\,L^r_{8}
          + 128\,L^r_{5}\,L^r_{6}
\extraline
          - 1024\,(L^r_{6})^2
          - 512\,L^r_{6}\,L^r_{8}
          - 32\,C^r_{13}
          - 16\,C^r_{15}
          + 32\,C^r_{20}
          + 192\,C^r_{21}
          + 32\,C^r_{32}
          ]
\extraline

       + \ln\myfrac{4}{3} \, [
          - 64/9\,L^r_{1}
          - 16/9\,L^r_{2}
          - 16/9\,L_{3}
          + 80/9\,L^r_{4}
          + 16/9\,L^r_{5}
          - 64/9\,L^r_{6}
\extraline
          - 32/9\,L^r_{8}
          ]

          - 176/9\,L^r_{1}
          - 124/27\,L_{3}
          + 64/3\,L^r_{4}
          + 140/27\,L^r_{5}
          - 208/9\,L^r_{6}
\extraline
          + 64/9\,L_{7}
          - 8\,L^r_{8}
$ \\[1ex]
4
& $
         N^{-1} \, [
            67/144
          + 11/24\,\rho_{1}
          ]

          - 5/36\,\ln\myfrac{4}{3}\,N^{-1} 

       + N \, [
            128\,(L^r_{4})^2
          + 64\,L^r_{4}\,L^r_{5}
\extraline
          - 256\,L^r_{4}\,L^r_{6}
          - 128\,L^r_{5}\,L^r_{6}
          + 16\,C^r_{15}
          ]

       + \ln\myfrac{4}{3} \, [

            64/9\,L^r_{1}
          + 16/9\,L^r_{2}
          + 16/9\,L_{3}
\extraline
          - 16/9\,L^r_{4}
          ]

       + 176/9\,L^r_{1}
          + 124/27\,L_{3}
          - 16/9\,L^r_{4}
          + 4\,L^r_{5}
          - 16\,L^r_{6}
          - 8\,L^r_{8}
$ \\[1ex]
5
& $
         N^{-1} \, [
          - 67/576
          + 7/64\,\rho_{1}
          ]

          + 5/96\,\ln\myfrac{4}{3}\,N^{-1} 

       + N \, [
          - 8\,C^r_{13}
          + 8\,C^r_{62}
          - 8\,C^r_{81}
          ]

$ \\[1ex]
6
& $
         N^{-1} \, [
          - 163/288
          - 1/16\,\rho_{1}
          ]

          + 1/24\,\ln\myfrac{4}{3}\,N^{-1} 

       + N \, [
            32\,C^r_{13}
          + 8\,C^r_{64}
          ]

$ \\[1ex]
7
& $
         N^{-1} \, [
          - 1937/576
          + 5/9\,\rho_1
          + 2/27\,\rho_2
          ]

       - 1/288

          + 25/18\,\ln\myfrac{4}{3}\,N^{-1}
\extraline
          - 22/81\,\ln^2\myfrac{4}{3}\,N^{-1}

       + N \, [
            1152\,(L_{7})^2
          - 1152\,L_{7}\,L^r_{4}
          + 2304\,L_{7}\,L^r_{6}
\extraline
          + 768\,L_{7}\,L^r_{8}
          + 32\,(L^r_{4})^2
          - 128\,L^r_{4}\,L^r_{6}
          - 384\,L^r_{4}\,L^r_{8}
          + 128\,(L^r_{6})^2
          + 768\,L^r_{6}\,L^r_{8}
\extraline
          + 128\,(L^r_{8})^2
          + 16\,C^r_{16}
          - 24\,C^r_{19}
          - 56\,C^r_{20}
          - 24\,C^r_{21}
          - 16\,C^r_{31}
          - 48\,C^r_{32}
\extraline
          - 48\,C^r_{33}
          ]

       + \ln\myfrac{4}{3} \, [

            64/9\,L^r_{1}
          + 16/9\,L^r_{2}
          + 16/9\,L_{3}
          - 16/3\,L^r_{4}
          - 80/9\,L^r_{5}
\extraline
          + 32/9\,L^r_{6}
          - 32/3\,L_{7}
          + 128/9\,L^r_{8}
          ]

          - 26/9\,L^r_{2}
          - 43/54\,L_{3}
          - 8\,L^r_{5}
%\extraline
          + 16\,L^r_{8}
$
\end{longtable}
\begin{longtable}{cl}
\hline\hline
& $
p_1
$ \\
\hline\\[-2ex]
\endhead
\\[-2ex]
\hline\hline
\caption[]{\rule{0cm}{2em}The polynomial $p_1$ defined in
  Eq.(\ref{eq:p0p1p2}) for $F$, $\Sigma$, $l_1^r\ldots l_6^r,\;l_7$.}
\label{tab:2}
\endfoot
$ F 
$ & $
         7/3

          - 1/3\,\ln\myfrac{4}{3}

       + N \, [
            416/9\,L^r_{1}
          + 104/9\,L^r_{2}
          + 122/9\,L_{3}
          - 68/3\,L^r_{4}
          ]

$\\[1ex]
$ \Sigma
$ & $
          - 2/81\,\ln\myfrac{4}{3} 

       + N \, [
            592/9\,L^r_{4}
          + 64/3\,L^r_{5}
          - 1312/9\,L^r_{6}
          + 64/3\,L_{7}
          - 320/9\,L^r_{8}
          ]

$ \\[1ex]
1
& $
            5/6\,N^{-1} 

          + 8\,L^r_{1}
          + 4\,L^r_{2}
          + 3\,L_{3}
          - 4\,L^r_{4}
$ \\[1ex]
2
& $
            13/24\,N^{-1} 

          - 8\,L^r_{2}
          - 2\,L_{3}
$ \\[1ex]
3
& $
          - 1501/648\,N^{-1} 

          + 11/162\,\ln\myfrac{4}{3}\,N^{-1} 

          - 352/9\,L^r_{1}
          - 88/9\,L^r_{2}
          - 106/9\,L_{3}
\extraline
          + 440/9\,L^r_{4}
          + 124/9\,L^r_{5}
          - 496/9\,L^r_{6}
          - 248/9\,L^r_{8}
$ \\[1ex]
4
& $
            37/18\,N^{-1} 

          - 7/18\,\ln\myfrac{4}{3}\,N^{-1} 

          + 352/9\,L^r_{1}
          + 88/9\,L^r_{2}
          + 106/9\,L_{3}
          - 88/9\,L^r_{4}
\extraline
          - 16\,L^r_{6}
          - 8\,L^r_{8}
$ \\[1ex]
5
& $
            17/48\,N^{-1} 

          - L^r_{9}
          - 2\,L^r_{10}
$ \\[1ex]
6
& $
          - 7/24\,N^{-1}

          + 2\,L_{3}
          + 2\,L^r_{9}
$ \\[1ex]
7
& $
            73/18\,N^{-1}

          + 101/162\,\ln\myfrac{4}{3}\,N^{-1}

          + 208/9\,L^r_{1}
          + 52/9\,L^r_{2}
          + 61/9\,L_{3}
\extraline
          - 52/3\,L^r_{4}
          - 224/9\,L^r_{5}
          + 104/9\,L^r_{6}
          + 184/3\,L_{7}
          + 632/9\,L^r_{8}
$
\end{longtable}
\newcommand{\extraLINE}{$\\ & & $}
\begin{longtable}{cll}
\hline\hline
& $
N\,p_2
$
\hspace*{0.92cm}& $
\ln(\mk^2/\Lambda^2)
$
\\
\hline\\[-2ex]
\endhead
\\[-2ex]
\hline\hline
\caption[]{\rule{0cm}{2em}The coefficient $p_2$ from
  Eq.(\ref{eq:p0p1p2}) as well as the logarithm defined in
  Eq.(\ref{eq:lnLamb}) for $F$, $\Sigma$, $l_1^r\ldots l_6^r,\;l_7$.}
\label{tab:4}
\endfoot
$ F
$ & $
       - 11/12\,N
$ & $
       - 14/11
       + 2/11\,\ln\myfrac{4}{3}
       - 13/11\,\Lb_{1}
       - 13/22\,\Lb_{2}
       - 244/33\,L_{3}\,N
\extraLINE
       + 17/22\,\Lb_{4}
$ \\[1ex]
$ \Sigma
$ & $
       - 28/81\,N
$ & $
         1/28\,\ln\myfrac{4}{3}
       - 333/56\,\Lb_{4}
       - 81/14\,\Lb_{5}
       + 451/56\,\Lb_{6}
\extraLINE
       - 216/7\,L_{7}\,N
       + 75/28\,\Lb_{8}
$ \\[1ex]
1
& $
       - 1/4
$ & $
       - 5/3
       - 6\,L_{3}\,N
       - 3/4\,\Lb_{1}
       - 3/4\,\Lb_{2}
       + 1/2\,\Lb_{4}
$ \\[1ex]
2
& $
         3/8
$ & $
         13/18
       - \Lb_{2}
       - 8/3\,L_{3}\,N
$ \\[1ex]
3
& $
         211/648
$ & $
       - 1501/422
       + 22/211\,\ln\myfrac{4}{3}
       - 594/211\,\Lb_{1}
       - 297/211\,\Lb_{2}
\extraLINE
       - 3816/211\,L_{3}\,N
       + 990/211\,\Lb_{4}
       + 837/211\,\Lb_{5}
       - 682/211\,\Lb_{6}
\extraLINE
       - 465/211\,\Lb_{8}
$ \\[1ex]
4
& $
       - 5/9
$ & $
       - 37/20
       + 7/20\,\ln\myfrac{4}{3}
       - 33/20\,\Lb_{1}
       - 33/40\,\Lb_{2}
       - 53/5\,L_{3}\,N
\extraLINE
       + 11/20\,\Lb_{4}
       + 11/20\,\Lb_{6}
       + 3/8\,\Lb_{8}
$ \\[1ex]
5
& $
       - 1/16
$ & $
       - 17/6
       + \Lb_{9}
       - 2\,\Lb_{10}
$ \\[1ex]
6
& $
       - 1/8
$ & $
         7/6
       - 8\,L_{3}\,N
       - \Lb_{9}
$ \\[1ex]
7
& $
        17/1296
$ & $

        2628/17
          + 404/17\,\ln\myfrac{4}{3}

          + 702/17\,\Lb_{1}
          + 351/17\,\Lb_{2}
\extraLINE
          + 4392/17\,L_{3}\,N
          - 702/17\,\Lb_{4}
          - 3024/17\,\Lb_{5}
          + 286/17\,\Lb_{6}
\extraLINE
          + 39744/17\,L_{7}\,N
          + 2370/17\,\Lb_{8}
$
\end{longtable}
{\it Acknowledgements.}

The  support of
Hans Bijnens and Gerhard Ecker was very essential for this work. 
Hans Bijnens  made available to us many of the
two--loop calculations performed by him and his collaborators, and even
calculated a particular one for our use.
Gerhard Ecker provided  us with a Form program which
simplified the evaluation of the two--loop functional considerably. We are
grateful for many useful discussions with Roberto Bonciani, Gilberto
Colangelo, S\'ebastien Descotes--Genon, Andrey I. Davydychev, Thorsten Ewerth,
Roland Kaiser, Heiri Leutwyler and Jan Stern, and thank Hans Bijnens, Gilberto
Colangelo and Gerhard Ecker for useful comments concerning the manuscript. The
calculations were performed partly with Form 3.1
\cite{form}. 
This work was supported  by the Swiss
National Science Foundation, by RTN, BBW-Contract No. 01.0357,
and EC-Contract  HPRN--CT2002--00311 (EURIDICE),
by the Ministerio de Educaci\'on y Ciencia under the project FPA2004-00996,
by Generalitat Valenciana GVACOMP2007-156, and by EU MRTN-CT-2006-035482
(FLAVIA{\it net}).

% --- REFERENCES -----------------------------------------------------------

\ed